\documentclass[english,useAMS,usenatbib]{mn2e}
\usepackage{lmodern}

\usepackage[T1]{fontenc}
\usepackage[latin9]{inputenc}
\usepackage{color}
\usepackage{amsmath}
\usepackage{amssymb}
\usepackage{graphicx}
\usepackage{esint}
\usepackage{placeins}
\usepackage[authoryear]{natbib}

\usepackage{hyperref}
\definecolor{darkgreen}{rgb}{0.1,0.6,0.7}
\hypersetup{colorlinks = true,urlcolor=blue,citecolor=darkgreen}

\makeatletter
\usepackage{lmodern}\usepackage{epstopdf}

\newcommand{\p}{\boldsymbol{p}}
\newcommand{\half}{\frac{1}{2}}

\newcommand{\Fab}{\ensuremath{F_{\alpha \beta}  }}
\newcommand{\vemu}[1]{\ensuremath{ \boldsymbol{\mu}_{#1} }}

\newcommand{\trace}[1]{\ensuremath{ \mathrm{Tr}\left[ {#1} \right]   }}
\newcommand{\logl}{\mathcal{L}}
\newcommand{\deltap}[1]{\Delta p_{#1}}
\newcommand{\tayC}{\ensuremath{\mathcal{T}}_{(C)}^{1}  }
\newcommand{\boX}{\ensuremath{\boldsymbol{x}}}
\newcommand{\bomu}{\ensuremath{\boldsymbol{\mu}}}
\newcommand{\boC}{\boldsymbol{C}}
\newcommand{\invboC}{\boldsymbol{C}_0^{-1}}
\newcommand{\bozero}{\boldsymbol{C_0}}

\newcommand{\dalilink}{\href{http://lnasellentin.github.io/DALI/}{DALI}}

\title[DALI: a non-Gaussian likelihood approximation]{A fast, always positive definite and normalizable approximation of non-Gaussian likelihoods}
\author[E. Sellentin]{Elena Sellentin$^{1}$\\
$^{1}$Institut f\"ur Theoretische Physik, Ruprecht-Karls-Universit\"at Heidelberg, Philosophenweg 16, 69120 Heidelberg, Germany\\
}

\makeatother

\usepackage{babel}
\begin{document}

\date{Accepted 2015 July 21. Received 2015 July 21; in original form 2015 May 27}

\maketitle
\pagerange{\pageref{firstpage}--\pageref{lastpage}} \pubyear{2015}

\label{firstpage} 
\begin{abstract}
In this paper we extent the previously published DALI-approximation for likelihoods to cases in which the parameter dependency is in the covariance matrix. 
The approximation recovers non-Gaussian likelihoods, and reduces to the Fisher matrix approach in the case of Gaussianity. It works with the minimal 
assumptions of having Gaussian errors on the data, and a covariance matrix that possesses a converging Taylor approximation.
The resulting approximation works in cases of severe parameter degeneracies and in cases where the Fisher matrix is singular. 
It is at least $1000$ times faster than a typical Monte Carlo Markov Chain run over the same parameter space. Two example applications, 
to cases of extremely non-Gaussian likelihoods, are presented -- one demonstrates how the method succeeds in reconstructing completely 
a ring-shaped likelihood. A public code is released here: \dalilink .
\end{abstract}

\section{Introduction}
Evaluating a multidimensional likelihood can be a computationally costly procedure. 
If speed matters, often a good approximation of the likelihood is required.
A widely used approximation of likelihoods is the Fisher matrix approximation, which singles out the Gaussian part of a likelihood 
\citep{Tegmark:1996bz}. Because many analytical results for Gaussians are available, such as the position of the 1-$\sigma$ confidence 
contours and higher-order equivalents, the Fisher matrix approximation is numerically fast to evaluate. 
It has also become widely used as it allows for the easy computation of Figures of Merit, simple determinants of the matrix elements and manipulations 
thereof, that can be used to evaluate the expected performance of an 
experiment, for example as introduduced to dark energy research by \citet{2006astro.ph..9591A}.

The alternatives to the Gaussian approximation are grid-evaluations of the likelihood, or sampling techniques such as Monte Carlo Markov Chains (MCMC), Nested Sampling \citep{2013JCAP...02..001A, 2014MNRAS.437.3918A,2004AIPC..735..395S}, and Population Monte Carlo (that uses iterative updates of a mixture 
model to capture non-Gaussianities \citep{2010MNRAS.405.2381K,2009PhRvD..80b3507W}). These methods tackle the challenge of characterising non-Gaussian likelihoods by using sophisticated algorithms. 
Gram-Charlier and Edgeworth-type expansions can also be used to capture non-Gaussianities, but suffer from regions in the parameter space 
where the approximated likelihood turns negative, thereby violating the Kolmogorov axioms for a probability \citep{CramerH}.

Nonetheless, likelihood approximations are urgently needed throughout the physical sciences, whenever evaluating a full 
likelihood is numerically too costly, e.g. when forecasting parameter constraints of a future experiment, where many different 
configurations need to be simulated, see e.g. \citep{2012MNRAS.422...44P,2011arXiv1110.3193L}. A quick check of the resulting 
likelihood is also desirable when optimizing a data analysis pipeline, or when establishing novel observables and testing how 
precisely they can constrain model parameters, see e.g. \citep{2014arXiv1409.3364C}. Non-Gaussian likelihood approximations, 
that maintain positive definiteness and normalizability, whilst rivaling the Fisher matrix in manners of speed, have recently 
become a focus of research. Transformations of the likelihood to Gaussianity are one way of tackling this problem \citep{Joachimi:2011iq}.
Another approach named `DALI' was presented in \citet{Sellentinetal} (henceforth named 'Paper1'), 
under the additional constraint of the data being Gaussianly distributed and the covariance matrix being constant. The main results of 
Paper 1 were application independent, i.e. the presented approximation would work for all observables to which it would be specified. 
The appendix contained insights into how the non-Gaussian likelihood approximation could also be applied to cases where the covariance matrix 
depends on parameters - however, additional assumptions about the covariance matrix needed to be made, which are fullfilled only for specific applications. 

In this paper, we extend the results of Paper 1 and present a non-Gaussian likelihood approximation that can deal with 
parameter-dependent covariance matrices. The main results will again be independent of the physical application, meaning the method 
can be applied in any field of physics, as well as in cosmology, or any other scientific branch that compares a parameterized model to data. 
The method only demands that the data shall be again Gaussianly distributed. Therefore, a public code \dalilink ~is being released 
along with this paper which allows the user to interface the DALI-formalism with their physical problems. The code also contains the 
results of Paper 1. A cosmological application to weak lensing will be presented in {\color{darkgreen}{Sellentin \& Sch\"aefer 2015}} (in prep.).
\section{Gaussianity}
Throughout the paper, we assume a data set $\boX$ with Gaussian errors, leading to the unapproximated likelihood
\begin{equation}
 L( \boX | \p) = \frac{1}{\sqrt{ (2\pi)^d |\boC|}}\exp\left(-\half (\boX - \vemu{})^T\boC^{-1} (\boX - \vemu{})      \right)
 \label{L}
\end{equation}
where $\p$ is a vector that holds $p$ parameters. The mean of the data $\bomu$ and the covariance matrix $\boC$ are predicted 
by a parameterized physical model and can in general both depend on the $p$ parameters. These parameters shall be constrained 
by maximizing the likelihood using the data which is collected in the data vector $ \boX $. The number of data 
points is $d$ and $|\boC|$ is the determinant of the covariance matrix. The covariance matrix is given by
\begin{equation}
 \boC(\boldsymbol{p}) = \langle \left(\boX - \bomu\right)\left(\boX - \bomu\right)^T\rangle,
\end{equation}
such that for a linear model, the parameters enter already quadratically in the covariance matrix. In general, the parameter 
dependence of the covariance matrix will be determined by the estimator applied and often also include nuisance parameters \citep{2014MNRAS.442.2728T}. 

The corresponding log-likelihood $\mathcal{L} = -\ln(L)$ of the Gaussian Eq.~(\ref{L}) is
\begin{equation}
 \mathcal{L} = \half \trace{\ln (\boC)  + \boldsymbol{C}^{-1} \langle \boldsymbol{D} \rangle},
 \label{Llog}
\end{equation}
where we neglected the $2\pi$ factors of the normalization, and where $\boldsymbol{D} = (\boX - \vemu{})(\boX -\vemu{})^T$ is the 
data matrix. Angular brackets denote averaging over the data.

The numerical costs of evaluating this likelihood will increase with the number of data points, the complexity of calculating the model 
predictions $\bomu$ and the estimation of the covariance matrix under variation of the parameters. In case of Bayesian inference, 
the likelihood could be updated to a posterior by multiplying with priors and normalizing by the corresponding evidence.

The assumption of Gaussian errors is not a severe constraint, since due to the central limit theorem, all data that 
stems from a distribution of finite variance, can be rebinned into a data set with Gaussian errors - if enough data points are available.
However, having Gaussian errors in the \emph{data} space does not mean that the resulting likelihood will be Gaussian in the \emph{parameter} space. 
Therefore, the mathematical tools available to exploit Gaussian likelihoods, such as their analytical marginalization over nuisance parameters, 
cannot be automatically exploited in the parameter space. The Gaussianity of the data set only transfers to the parameter space if 
no parameter degeneracies occur and if the model that is compared to the data is linear in all parameters. Similarly, a Gaussian likelihood can 
also be expected if the data set is constraining enough, such that essentially a linear Taylor approximation of the model and the covariance 
matrix around the best fit point is sufficient. This explains why the Fisher matrix has become so popular in forecasting the performance 
of precision experiments, which were designed to tightly constrain targeted parameters. 

In contrast, achieving extremely constraining 
data with a new experiment cannot be expected by default if for example extensions to a standard model are to be investigated and new parameters 
measured for the very first time. If the forecasted data is not expected to be extremely constraining, the likelihood will not be peaked so 
sharply around the best fit that a linear Taylor approximation of the model, and the covariance matrix alone may not be good enough. 
This already hints at why the following non-Gaussian likelihood approximation needs to build on higher order derivatives.

The higher order likelihood approximation for a constant covariance matrix was derived in Paper 1.
Here, we specialize to the case of the model dependence of the mean being identically zero, $\bomu(\boldsymbol{p}) \equiv 0$, 
and all parameter dependence is contained in the covariance matrix.
This can be the case in a real scenario, where the mean is zero but fluctuations around that mean can be of different amplitudes, and 
this is encoded in the covariance. Examples are a measurement of pure noise, which clearly has mean zero, but where the covariance of the noise 
depends on parameters. Another example is any kind of mode decomposition, where again it is clear that a mode has mean zero. 
A cosmological example is the galaxy power spectrum, which arises from density fluctuations around the cosmic mean value, 
and where the mean overdensity must be zero, due to mass conservation. The power spectrum can then be used as the covariance in the following framework, where it is the covariance of the Fourier amplitudes of the overdensity field.\footnote{Often, however, such analyses 
are carried out by comparing a measured power spectrum to a parameterized power spectrum, which is then treated as the mean. In these cases the 
covariance matrix would then be the covariance \emph{of} the power spectrum (a four-point function) instead of \emph{the} powerspectrum.} 

\section{Problems when approximating likelihoods}
\label{problems}
Approximating a likelihood is more complicated than approximating a more general function because one typically wishes the likelihood to 
be positive semi-definite at all orders; otherwise negative probabilities occur, which are non-sensical. 
Positive semi-definiteness is a strong constraint and not automatically fulfilled by a usual Taylor series approximation of the likelihood. 
For example, Taylor approximating a standard normal distribution yields,
\begin{equation}
\exp(-x^2) = 1 -x^2 + \frac{1}{2}x^4 + \mathcal{O}(x^5).
\label{TaylorL}
\end{equation}
If truncated at second order, this approximation becomes negative at 2-$\sigma$ from the best fit, or begins rising to 
infinity at about 2-$\sigma$ when truncated at fourth order. This divergence makes the likelihood approximation not 
normalizable, such that no measure for relative likelihoods can be defined. Both, second and fourth order approximation of 
the standard normal distribution therefore violate defining properties of a likelihood. Obviously, a continuation of 
the Taylor approximation Eq.~(\ref{TaylorL}) to very high orders would remedy both of these issues but this would be a cumbersome approach. It 
is well known that Taylor approximating the log-likelihood instead, reconstructs the Gaussian likelihood much more quickly
\begin{equation}
 \exp(-x^2) = \exp(-\mathcal{L}) = \exp(-\mathcal{T}(\mathcal{L})),
\label{TayLog}
\end{equation}
where $\mathcal{T}(\mathcal{L})$ denotes the Taylor series of the log-likelihood. If this Taylor series is evaluated at the maximum of the standard 
normal distribution then already the first and second order terms of this series recovers the Gaussian likelihood completely, and 
all higher orders of the series are identically zero. The approximation schemes Eq.~(\ref{TaylorL}) and Eq.~(\ref{TayLog}) are 
both mathematically valid ways of approximating the standard normal distribution, even though they lead to entirely different Taylor series.
The scheme outlined in Eq.~(\ref{TayLog}) is however much more advantageous because it leads already at second order the desired approximation, 
and negative likelihoods then do not appear at all, since the exponential function is always positive. Therefore, the choice of which quantity shall 
be approximated influences decisively how quickly the approximation recovers the shape of the original function, and whether unwanted artifacts 
appear when truncating the approximation at low orders.

The choice of Taylor approximating the log-likelihood, instead of the likelihood, to second order in multiple dimensions 
yields a Hessian matrix whose expectation value 
is the Fisher (or Information) matrix. Denoting partial derivatives by $\partial_\alpha f = f,_\alpha$, the Fisher matrix of Eq.~(\ref{L}) can be written as
\begin{equation}
 \begin{aligned}
 \Fab = &  \langle \logl,_{\alpha\beta} \rangle |_{\boldsymbol{\hat{p}}} \\
 = & \frac{1}{2} \trace{\boldsymbol{C_0}^{-1} C,_\alpha \boldsymbol{C_0}^{-1} C,_\beta\ } + \bomu,_\alpha \boldsymbol{C_0}^{-1} \bomu,_\beta
 \end{aligned}
  \label{fish_formula}
\end{equation}
where the derivatives are evaluated at the maximum likelihood point $\boldsymbol{\hat{p}}$ and summation over repeated indices is implied. 
All quantities that are to be evaluated at the maximum likelihood point are marked by a subscript zero. Consequently, $\bozero$ is constant and cannot be 
derived with respect to parameters.

The corresponding likelihood approximation is then given by
\begin{equation}
L(\boX | \p ) \approx N \cdot \exp( -\half \Fab \deltap{\alpha}\deltap{\beta})
\label{Lfish}
\end{equation}
where the $\deltap{\alpha} =  p_\alpha - \hat{p}_\alpha$ are the offsets from the best fit point $\hat{p}_\alpha$ and $N$ is a normalization constant.

The Fisher approximation results in the usual ellipsoidal, multi-variate correlated Gaussian confidence contours, 
which often do not recover the shape of a non-Gaussian likelihood distribution well. A continuation of the Taylor series is 
then desirable in order to capture these non-Gaussianites. This wish for a continuation of the series is predicated on the requirement 
to solve the issue of normalizability and positive-definiteness at all orders. 
Also, it is preferrable to recover the essential shape of the likelihood with as little additional terms as possible for computational efficiency.
Clearly, just as there exist multiple ways in approximating the likelihood Eq.~(\ref{TaylorL}), there will exist multiple ways of 
continuing the approximation from that given by the Fisher matrix. 
These extended approximations will pick up the desired information about the likelihood's shape with different efficiencies.
An obvious extension would be the continuation of the log-likelihood's Taylor-approximation
\begin{equation}
\begin{aligned}
L(\boX | \p) \approx & N \exp \left(  -\half \Fab \deltap{\alpha}\deltap{\beta} \right.  \\
& -\frac{1}{3!} S_{\alpha\beta\gamma}\deltap{\alpha}\deltap{\beta}\deltap{\gamma} \\
& \left.  -\frac{1}{4!} Q_{\alpha\beta\gamma\delta}\deltap{\alpha}\deltap{\beta}\deltap{\gamma}\deltap{\delta} + \mathcal{O}(\deltap{}^5)\right),  
\label{taylorlog}
\end{aligned}
\end{equation}
where
\begin{equation}
\begin{aligned}
 S_{\alpha\beta\gamma} = & \logl,_{\alpha\beta\gamma}|_{\hat{\p }}\\
 & = -2\trace{\invboC \boC,_\gamma \invboC \boC,_\beta \invboC \boC,_\alpha } \\
 & + \frac{3}{2} \trace{ \invboC \boC,_\gamma \invboC \boC,_{\alpha \beta}},\\
 \end{aligned}
 \label{flex}
\end{equation}
and
\begin{equation}
 \begin{aligned}
Q_{\alpha \beta\gamma\delta}  = & \logl,_{\alpha\beta\gamma\delta}|_{\boldsymbol{\hat{p}} }\\
& = 9\ \trace{\invboC \boC,_\delta \invboC \boC,_\gamma \invboC \boC,_\beta \invboC \boC,_\alpha }\\
& + \frac{3}{2} \trace{\invboC \boC,_{\gamma\delta} \invboC \boC,_{\alpha \beta}} \\ 
& - 12\ \trace{ \invboC \boC,_{\gamma\delta}\invboC \boC,_\beta \invboC \boC,_\alpha }\\
& + 2\ \trace{\invboC \boC,_\gamma \invboC \boC,_{\alpha\beta\delta}}\\
 \end{aligned}
 \label{quarx}
\end{equation}
which gives the Taylor series of the log-likelihood up to fourth order, after being averaged over the data. 

In reference to Eq.~(\ref{taylorlog}) multiple observations can be made. Firstly, this approximation will in general 
be unnormalizable since it will diverge somewhere in parameter space. This is partly due to the odd powers of $\deltap{}$, which 
will clearly become negative on one side of the fiducial point (about which the expansion is made) if they are positive on the other side 
of the fiducial point; the argument of the exponential function will then become positive 
even for small displacements from the best fit point, and the approximation will begin to diverge. Also 
the summation over even powers of $\deltap{}$ can lead to divergences, as terms of the structure $\deltap{1}\deltap{1}\deltap{1}\deltap{2}$ 
will appear, as has been detailed in Paper 1. 
These divergences of the approximation can only be avoided in an application-independent way if the argument of the exponential is negative everywhere in parameterspace. One way of achieving this is to demand the approximation to have the shape
\begin{equation}
 L \approx N \exp\left( - Q \right),
\label{quad}
\end{equation}
where $Q$ is a quadratic function of the parameters and therefore always positive definite. 
The expansion Eq.~(\ref{taylorlog}) of the log-likelihood does not have this shape.

Secondly, we observe that even if only first order derivatives of the covariance matrix were non-vanishing, the above series would 
still not terminate after the Fisher approximation. The first lines of Eq.~(\ref{flex}) and Eq.~(\ref{quarx}) contain only 
first order derivatives of the covariance matrix and make it clear that at the $n$-th Taylor order a term of the shape
\begin{equation}
\trace{\left(\bozero^{-1}\boC,_\alpha \deltap{\alpha}\right)^n}
\label{highT}
\end{equation}
appears, where we have expressed the repeated multiplication of the same matrices as a power. As new information on the parameter dependence 
of the covariance matrix is encoded in its higher order derivatives, the terms Eq.~(\ref{highT}) do not add any of the 
new information which we target; they simply stem from the slowly convergent Taylor series of the logarithm.

Therefore we see that a Taylor approximation of the log-likelihood beyond second order is a valid but 
laborious way to include non-Gaussian behaviour: the log-likelihood would need to be approximated to much higher than the 4th order, 
before it can be expected to be normalizable for a physical application.
To avoid all of the above discussed difficulties, we construct a likelihood approximation in which we explicitely request 
that it shall have the shape Eq.~(\ref{quad}). This can be achieved by Taylor approximating the covariance matrix directly 
which has the further advantage that $\boC$ depends on the parameters more sensitively than $\ln(\boC)$. Thus, Taylor expanding  
the covariance matrix will pick up the higher order derivatives earlier.
This approximation is deduced in Sec.(\ref{beyond}) and the convergence criterion for this approximation is given in Sec.(\ref{sec:appl})

\section{Beyond Gaussianity}
\label{beyond}
We express the variation of the covariance matrix over the parameter space by its Taylor series and single out the constant zeroth-order term
\begin{equation}
 \boC(\boldsymbol{p}) = \boldsymbol{C_0} + \tayC,
\end{equation}
where $\bozero$ is the constant covariance matrix evaluated at the likelihood maximum, and
\begin{equation}
 \tayC = \sum_{n=1}^\infty \frac{\boC^{(n)}|_{\boldsymbol{\hat{p}}}}{n!} (p_\alpha -\hat{p}_\alpha)...(p_n -\hat{p}_n)
\end{equation}
is the $p$-dimensional Taylor series of the covariance matrix, beginning at the first derivative $\boC^{1}$. The derivatives are 
chosen to be evaluated at the maximum of the likelihood, denoted by $\boldsymbol{\hat{p}}$. This series carries information on how 
the covariance matrix changes throughout the parameter space. Here, we are specifically interested in higher order derivatives of 
the covariance matrix, since these encode the non-linear dependence of the covariance matrix on parameters.
For $\bomu \equiv 0$ the data matrix is $\boldsymbol{D} = \boX \boX^T$ which is parameter independent. The log-likelihood is then given by
\begin{equation}
\begin{aligned}
 \mathcal{L} & = \half \trace{ \ln(\boC) } + \half \trace{ \langle \boX \boX^T \rangle \boC^{-1}}\\
 & = \half \trace{ \ln\left(\bozero[1+\bozero^{-1} \tayC  ] \right) + \langle \boX \boX^T \rangle \left( \bozero+\tayC  \right)^{-1}    },
\label{logex}
\end{aligned}
\end{equation}
where angular brackets denote averaging over the data and $\langle \boX \boX^T \rangle$ is kept explicitely, 
in order to emphasize that it does not depend on parameters, although it will later average out to be the measured covariance matrix. 
So far, the covariance matrix has only been rewritten, but no approximation has been made.

However, if the Taylor series $\tayC$ is evaluated only sufficiently close to the maximum likelihood point, then 
we will have $\tayC \ll \bozero$ and we can consistently approximate Eq.~(\ref{logex}) up to second order in $\tayC$. 
This leads to the targeted shape Eq.~(\ref{quad}).
We therefore approximate by applying the matrix inversion identity (also known as Woodbury identity)
\begin{equation}
 (\boldsymbol{A} +\boldsymbol{B} )^{-1} = \boldsymbol{A}^{-1} - \boldsymbol{A}^{-1}\left( 1 + \boldsymbol{B}\boldsymbol{A}^{-1}\right)^{-1}\boldsymbol{B}\boldsymbol{A}^{-1}
\end{equation}
to find an approximation for the inverted covariance matrix
\begin{equation}
\begin{aligned}
 \left( \bozero+\tayC  \right)^{-1} & = \invboC + \invboC \tayC\ \invboC \\
 & - \invboC \tayC\ \invboC \tayC\ \invboC + \mathcal{O}(3),
\label{inv}
 \end{aligned}
\end{equation}
where the approximation was truncated at second order since we target the shape Eq.~(\ref{quad}). 
The quadratic term of the logarithm's Taylor expansion is,
\begin{equation}
 \ln (1+x) = x -\frac{x^2}{2} + \mathcal{O}(x^3).
\label{log}
\end{equation}
The quadratic approximation of the log-likelihood then becomes
\begin{equation}
\begin{aligned}
 \mathcal{L} &\approx \half \trace{\ln(\bozero)  + \invboC \tayC - \half \invboC \tayC\  \invboC \tayC }\\
 &+ \half \mathrm{Tr}\left[ \langle \boX \boX^T \rangle \left( \invboC - \invboC \tayC \invboC \right. \right. \\
 &\ \ \ \ \ \ \ \ \ \ \  \ \ \ \ \ \ \ \ \ \ \ \ + \left. \left. \invboC \tayC \invboC \tayC \invboC  + \mathcal{O}(3) \right) \right].
\label{puttogehter}
 \end{aligned}
\end{equation}
Applying $\langle \boX \boX^T \rangle = \bozero$ the likelihood approximation simplifies to
\begin{equation}
\begin{aligned}
L & \approx N \exp( -\mathcal{L} ) \\
  & = N \exp \left( - \frac{1}{4} \trace{ \invboC \tayC \invboC \tayC } + \mathcal{O}(3) \right)\\
&=  N \exp \left( - \frac{1}{4} \mathrm{Tr}\left[ \invboC (\boC,_\alpha \deltap{\alpha} +\half \boC,_{\alpha\beta} \deltap{\alpha}\deltap{\beta} + ...) \right. \right.\\
& \left. \left. \ \ \ \ \ \ \ \ \ \  \ \ \ \ \ \ \ \ \ \ \ \ \ \ \ \ \invboC (\boC,_\alpha \deltap{\alpha} +\half \boC,_{\alpha\beta} \deltap{\alpha}\deltap{\beta} + ...) \right] + \mathcal{O}(3) \right),
 \end{aligned}
\end{equation}
where $\ln(\bozero )$ and $\bozero \invboC = 1$ are constants and were absorbed into the normalization constant $N$. In the last step, a repeated multiplication of the same terms appears. This can be rewritten as
\begin{equation}
\begin{aligned}
L & \approx\\
  &= N \exp \left( - \frac{1}{4} \trace{ \left(\invboC (\boC,_\alpha \deltap{\alpha} +\half \boC,_{\alpha\beta} \deltap{\alpha}\deltap{\beta} + ...) \right)^2 } \right),
\label{mainres}
 \end{aligned}
\end{equation}
where the repeated multiplication of the same matrices in the trace was made more explicit by denoting it as a square. 

We therefore have arrived at an approximation of the shape Eq.~(\ref{quad}) that includes higher order derivatives of the covariance matrix. 
This approximation will consequently remain normalizable everywhere in parameter space.
This result generalizes the usual Fisher matrix in a straight forward way: if the Taylor-approximation $\tayC$ is truncated 
at first order, the usual Fisher matrix approximation Eq.~(\ref{fish_formula}) of the likelihood is obtained and the higher order corrections are then
\begin{equation}
\begin{aligned}
 L \approx  N \exp & \left(  - \frac{1}{4} \trace{ \invboC \boC,_\alpha \invboC \boC,_\beta }\deltap{\alpha}\deltap{\beta} \right. \\
& \left. - \frac{1}{4} \trace{ \invboC \boC,_\alpha \invboC \boC,_{\beta\gamma} } \deltap{\alpha}\deltap{\beta}\deltap{\gamma} \right.\\
& \left. - \frac{1}{16} \trace{ \invboC \boC,_{\alpha\beta} \invboC \boC,_{\gamma\delta} } \deltap{\alpha}\deltap{\beta}\deltap{\gamma}\deltap{\delta} \right.\\
& \left. - \frac{1}{24} \trace{ \invboC \boC,_{\alpha\beta} \invboC \boC,_{\gamma\delta\epsilon} } \deltap{\alpha}\deltap{\beta}\deltap{\gamma}\deltap{\delta}\deltap{\epsilon} \right.\\
& \left. - \frac{1}{144} \trace{ \invboC \boC,_{\alpha\beta\gamma} \invboC \boC,_{\delta\epsilon\phi} } \deltap{\alpha}\deltap{\beta}\deltap{\gamma}\deltap{\delta}\deltap{\epsilon}\deltap{\phi} \right),
 \end{aligned}
\end{equation}
where we have chosen to truncate the Taylor expansion of the covariance matrix at third order for brevity;  
the continuation to fourth and higher orders of the covariance matrix is however obvious from Eq.~(\ref{mainres}).
The terms that are cubic and quintic in the $\deltap{}$ can become negative and thereby decrease the likelihood estimate in regions, 
where it had been overerstimated by the even-order terms. In total however, the terms combine to a quadratic form, 
and thereby the approximation is known to not diverge anywhere in parameter space.
 
As this result generalizes the findings of Paper 1, and is also based on a derivative expansion (this time of the covariance matrix), 
we stick with the name DALI (Derivative Approximation for LIkelihoods). If this approximate likelihood shall be updated to a posterior distribution, 
multiplication by a prior can be achieved by adding the log-likelihood of the prior to the DALI-approximation, just as in case of the 
Fisher matrix approximation. Details about the expected speed-up when compared to MCMC can be found in Paper 1.
\begin{figure*}
\includegraphics[width=\textwidth]{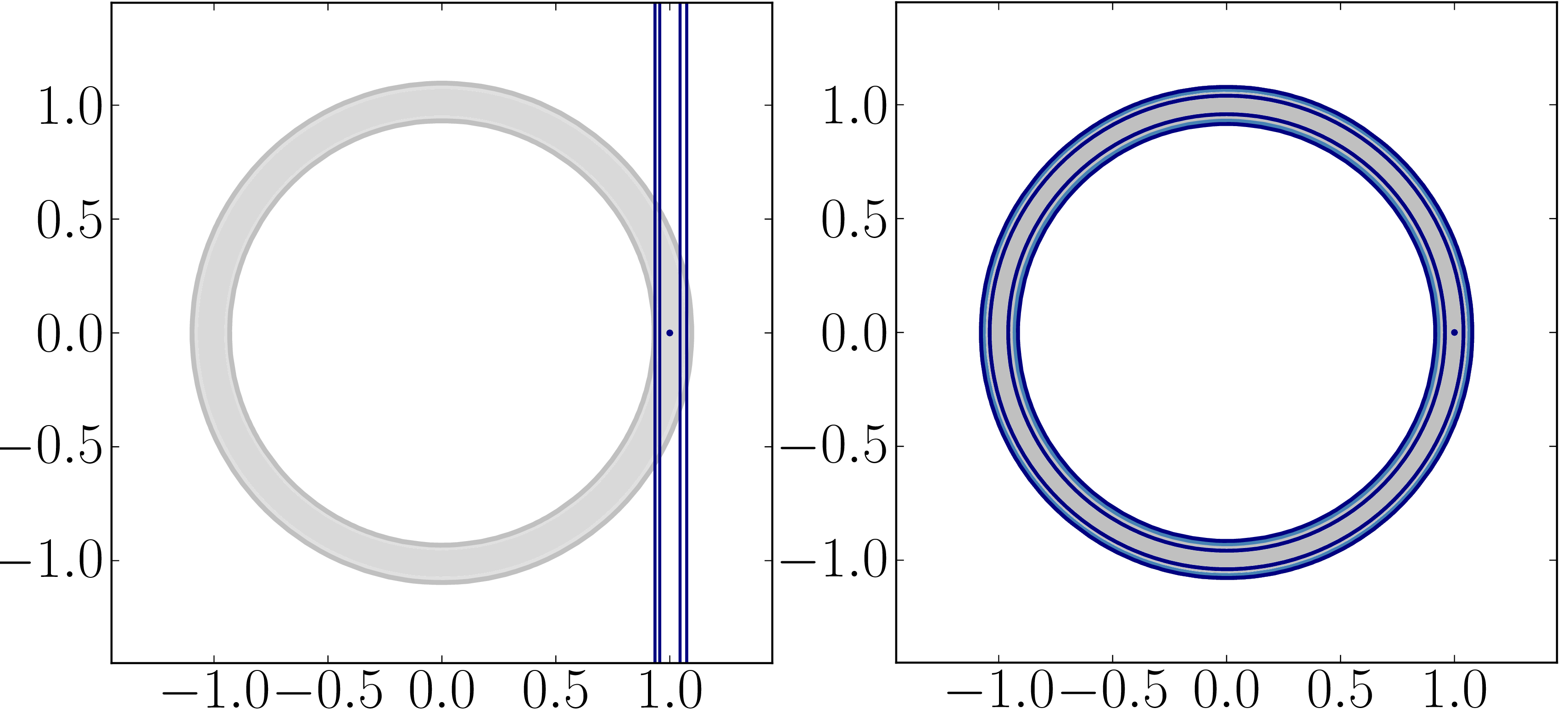} 
\caption{The unapproximated likelihood of Eq.~(\ref{mod1}) is depicted in grey. Since Eq.~(\ref{mod1}) is the equation of a circle, the likelihood has a ring-shape. Left: The Fisher approximation in blue, with fiducial point indicated by the small blue dot. The Fisher matrix is singular and therefore apprears as a set of parallel lines. Right: in blue the DALI approximation using second order derivatives of the covariance matrix Eq.~(\ref{mod1}).  }
\label{circle_fish} 
\end{figure*}

\begin{figure*}
\includegraphics[width=\textwidth]{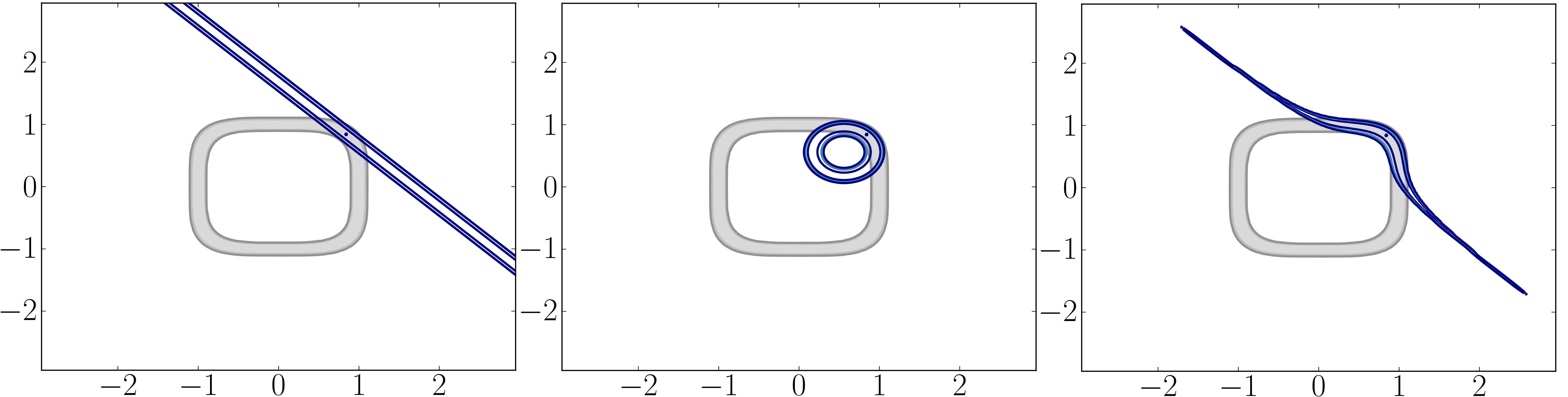} 
\caption{Like Fig.(\ref{circle_fish}) only for a likelihood using the covariance matrix Eq.~(\ref{mod2}). The likelihood is again depicted by the empty grey ring, and the different approximations are depicted in blue: Fisher matrix (left), DALI with second order derivatives of the covariance matrix (middle), DALI with third order derivatives (right).}
\label{square_2nd} 
\end{figure*}

\section{Criteria of applicability}
\label{sec:appl}
Non-Gaussianity can arise from at least two sources. For example if the data has only little constraining power then even the likelihood for a model with  
only mildly non-linear parameters will pick up non-Gaussianities. 
In contrast, if the data is very constraining non-Gaussianites will still occur if parameters are degenerate with each other over a finite range. 
In this case the non-Gaussianities can be recovered by DALI.

The approximation of Eq.~(\ref{mainres}) is strictly valid if the following criteria are fulfilled:
\begin{itemize}
 \item The data set $\boX$ must be so constraining that the likelihood is confined to within a region $\deltap{}$ where 
the second order Taylor approximations Eq.~(\ref{log}) and Eq.~(\ref{inv}) dominate over their higher orders.
\item Approximating the log in Eq.~(\ref{puttogehter}) requires 
\begin{equation}
\trace{ \invboC \tayC} \ll 1
\end{equation}
which can be solved for parameter offsets $\deltap{}$ 
\begin{equation}
 \deltap{\alpha} \ll \frac{1}{\trace{\invboC(\boC,_\alpha + \frac{1}{2}\boC,_{\alpha\beta}\deltap{\beta} + ...)}}
\label{cond}
\end{equation}
\end{itemize}
The last requirement will be fulfilled if the data set confines the preferred parameter space to an area within which 
the Taylor-approximation captures well the variation of the covariance matrix throughout the parameter space. DALI is therefore expected 
to work well in case of rather constraining data and degenerate parameters, while a good recovery of non-Gaussianities for weakly 
constraining data and mild non-linear dependences on the parameters would require Taylor-approximating the log-likelihood to much higher 
orders with the corresponding difficulties detailed in Sec.(\ref{problems}).
If the condition Eq.~(\ref{cond}) is only marginally fulfilled, the DALI-approximations will still converge although they will not pick up all the  
shape-information of the likelihood. Mismatches between the shape of the approximation and the real likelihood shape will then be observed. This is already known from the Fisher matrix, and expected to be more mild in DALI since the higher order derivatives will correct upon the Fisher matrix misestimates.

\section{Parameter dependent covariance matrix and mean}
The DALI formalism described above is able to recover non-Gaussian likelihood shapes if the covariance matrix depends on parameters. 
In the previous paper \citet{Sellentinetal}, a non-Gaussian likelihood approximation was developed for the case when the covariance matrix is constant, 
and only the mean $\bomu$ depends on parameters. An interesting question is whether the two approxiations can be combined to 
approximate a likelihood where \emph{both} mean and covariance matrix depend on data. Multiple interesting aspects should be pointed out in this context. One expects a likelihood approximation to fulfill three criteria: it shall be positive-definite, normalizable, and additionally possess a high degree of shape fidelity, i.e. quickly converge towards the shape of the unapproximated likelihood. Positive definiteness and normalizability are guaranteed by DALI being of the shape $L \approx \exp(-Q)$, with $Q$ being a positive definite form in the parameters. In principle, any positive definite form could be chosen. However, our choice of using the squared Taylor series of either $\bomu$ or $\boC$ additionally guarantees the shape fidelity of the DALI expansion. If the squared Taylor series were replaced by another quadratic form, the shape fidelity would most likely be quickly lost. 
If both, $\bomu$ and $\boC$ depend on the same parameters, no quadratic form that is at the same time a Taylor series has been found so far due 
to the appearance of crossterms between derivatives of $\bomu$ and $\boC$, e.g. $\bomu,_\alpha \boC,_\beta \bomu,_\gamma$. Neglecting these crossterms will produce a DALI-expansion that is a simple multiplication of Eq.~(\ref{mainres}) and Eq.~(16) from \citet{Sellentinetal}. This may be a good approximation in many cases, e.g. when the covariance matrix depends strongly on some parameters but not on those on which the mean depends. However, due to omitting the crossterms, in general this expansion will not be able to recover all information and therefore it may not yield a good approximation.

\section{Testcases}
The strength of this new approximation scheme was tested on two toy-models of particularly severe non-Gaussianities which arise from degeneracies.
Both toy models are two-dimensional and have $\bomu \equiv 0$. The data set consists of 50 data points. The covariance matrix 
of the first is diagonal and given by
\begin{equation}
 C_{ij}(\boldsymbol{p}) = (p_1^2 + p_2^2)\delta_{ij},
\label{mod1}
\end{equation}
with the Kronecker-Delta $\delta_{ij}$. Since $p_1^2 + p_2^2 = 1$ is the equation of a circle, this model produces a ring-shaped unapproximated 
likelihood, with the interior of the ring being a region of zero likelihood. All points which lie exactly on the circle will maximize the likelihood and any of them could be chosen as fiducial point for evaluating the DALI approximation. Taking more than 50 data points would decrease the thickness of the ring but would never be able to lift the degeneracy, even for an infinite number of measurements. Such likelihoods appear for example in particle physics for measurements of the Cabibbo-Kobayashi-Maskawa matrix \citep{2015PhRvD..91g3007C}.

The covariance matrix of the second toy model is,
\begin{equation}
  C_{ij}(\boldsymbol{p}) = (p_1^4 + p_2^4)\delta_{ij},
\label{mod2}
\end{equation}
which again possesses a closed degeneracy line of a somewhat boxy ring-shape. Again, each point along the line $p_1^4 + p_2^4 =1$ can serve as fiducial point for the DALI approximation. The unapproximated likelihoods of these two models are depicted as grey shades in Fig.(\ref{circle_fish}-\ref{square_2nd}), where the two shades indicate the 68$\%$ and $95\%$ confidence contours. Both toy models were then approximated by Eq.~(\ref{mainres}), truncated at different orders.
The Fisher matrix of both cases is degenerate and appears as parallel non-closing lines. Changing the evaluation of the derivatives 
cannot break this degeneracy. The second-order DALI-approximation already finds the full circle, since no higher than second order derivatives exist in this case. For the second toy-model, a complete recovery of the likelihood would require the calculation of fourth order derivatives. Although this could be done analytically in the case at hand, in general such a calculation would need a numerical solution. We therefore maintain the truncation of the expansion Eq.~(\ref{mainres}) at third order, as implemented in the public code \dalilink. The resulting approximation can be seen in Fig.~(\ref{square_2nd},~right) for third order derivatives, or second order derivatives in Fig.~(\ref{square_2nd},~middle). The degeneracy of the Fisher matrix is lifted in both cases, and the improvement in shape fidelity can easily be seen.
As typical applications of this method would not posess such strong parameter degeneracies, it can be expected that the DALI-method will reconstruct the likelihood contours with great accuracy. 

The DALI-code that combines the specialized likelihood approximations of our previous paper, and the extension presented here, is public at \dalilink. However, due to the structural similarity with the Fisher matrix, any already existing Fisher code can easily be upgraded to a DALI-code by adding the higher order derivatives of Eq.~(\ref{mainres}).

\section*{Acknowledgements}
It is a pleasure to thank Luca Amendola, Matthias Bartelmann, Alan Heavens, Tom Kitching and Bj\"orn Malte Sch\"afer for support and helpful discussions. This work has received financial support through the RTG \emph{Particle Physics beyond the Standard Model} (DFG fund 1904), and the transregional collaborative research centre TR 33 '\emph{The Dark Universe}' of the German Science Foundation.

\bibliographystyle{mn2e}
\bibliography{Dali2.bib}

\label{lastpage} 
\bsp 
\end{document}